# Colossal dielectric constants: A common phenomenon in $CaCu_3Ti_4O_{12}$ related materials

J. Sebald[a], S. Krohns[a], P. Lunkenheimer[a,*], S. G. Ebbinghaus[b], S. Riegg[c], A. Reller[c] and A. Loidl[a]

[a] *Experimental Physics V, Center for Electronic Correlations and Magnetism, University of Augsburg, 86135 Augsburg, Germany*
[b] *Solid State Chemistry, Martin-Luther University Halle-Wittenberg, 06120 Halle, Germany*
[c] *Solid State Chemistry, University of Augsburg, 86135 Augsburg, Germany*

[*]Corresponding author. *E-mail address:* peter.lunkenheimer@physik.uni-augsburg.de

**Abstract**
In the present work we demonstrate that in addition to the well-known colossal-dielectric-constant material $CaCu_3Ti_4O_{12}$ also various members of the series $Ln_{2/3}Cu_3Ti_4O_{12}$ with $Ln$ = La, Ce, Pr, Nd, Sm, Eu, Gd, Tb, Dy, Ho, Er and Tm, exhibit giant values of the dielectric constant. Just as $CaCu_3Ti_4O_{12}$, all these materials show a Maxwell-Wagner type relaxation process. For the best material, $Pr_{2/3}Cu_3Ti_4O_{12}$, we provide a detailed investigation of its dielectric properties in a broad frequency range up to 1 GHz. Polarization at internal barriers, most likely grain boundaries, seems to be the reason for the observed very high values of the dielectric constant. Taking into account the present results and those reported in literature, we conclude that $CaCu_3Ti_4O_{12}$ is not exceptional at all and there seem to be various isostructural materials with similar dielectric properties.

PACS: 77.22.Ch; 77.84.Dy

## 1. Introduction

$CaCu_3Ti_4O_{12}$ (CCTO) shows an extremely high ("colossal") dielectric constant (CDC), which, in contrast to ferroelectrics like $BaTiO_3$, is nearly constant in a broad temperature range [1,2,3,4,5,6,7]. Thus, this material has attracted tremendous attention due to its possible applications, e.g., for enhancing the performance of capacitive electronic elements. Numerous investigations have revealed that the dielectric properties of CCTO strongly depend on the processing conditions during preparation, the type of external contacts and the measuring frequency. It is more or less commonly accepted that the CDC of CCTO is caused by an extrinsic mechanism, arising, e.g., from an internal barrier layer capacitance (IBLC) [3,4,8,9,10] or a surface barrier layer capacitance (SBLC) [6,7,11,12,13]. However, from an application point of view, this is rather irrelevant and we would like to recall that interface effects are the basis of the high capacitance of commercially available double-layer capacitors [14].

There is a large number of compounds, isostructural to CCTO [2,4,15,16,17]. Is CCTO really the only one with outstanding dielectric properties? The interest just in CCTO was originally triggered by a list of thirteen $ACu_3B_4O_{12}$ compounds, published in 2000 by Subramanian *et al.* [2], which provided the room temperature values of the dielectric constant measured at 100 kHz. In that list, which was later complemented by ten additional materials [4], CCTO showed the highest dielectric constant of all investigated compounds. However, as mentioned above, the absolute values of the CDCs in CCTO are strongly dependent on various factors, like processing conditions during sample preparation and frequency. Thus measurements at a single frequency in one polycrystalline sample may not be sufficient to exclude the occurrence of CDCs in other materials and it may be possible to find similarly spectacular dielectric properties also in other CCTO-related compounds. Indeed CDCs were reported for some of those materials [4,15,18]. To further elucidate this issue, we prepared a number of polycrystalline $Ln_{2/3}Cu_3Ti_4O_{12}$ (LnCTO) samples with $Ln$ = La, Ce, Pr, Nd, Sm, Eu, Gd, Tb, Dy, Ho, Er and Tm and screened them with dielectric spectroscopy up to 1 MHz. (For Ce, evidence for a mixed valence state between 3+ and 4+ was found and the actual composition may be $Ce_{1/2}Cu_3Ti_4O_{12}$. This issue will be clarified by further investigations.) As an example of a material with dielectric properties comparable to those of CCTO, we focused on $Pr_{2/3}Cu_3Ti_4O_{12}$ (PCTO), which was investigated in a broader frequency range up to 1 GHz.

## 2. Experimental details

Polycrystalline samples of LnCTO were prepared by solid state reaction. The binary oxides $Ln_2O_3$ (exceptions: $Pr_6O_{11}$, $CeO_2$ and $Tb_4O_7$), CuO and $TiO_2$ were mixed in corresponding molar ratios and well ground in a mortar. Before weighting, all lanthanide oxides were dried at 900°C for 12 h to remove any water in the samples. In the same way $Pr_2O_3$ was reacted to non-hygroscopic $Pr_6O_{11}$. An excess of 0.3 g - 0.4 g CuO was added as flux material. After completion of the reaction, this excess was removed by washing the samples with hydrochloric acid (10%) and afterwards with deionized water. Before calcination, the samples were pressed into pellets. Calcination in aluminumoxide crucibles was done in two steps, first at 1000°C for 48 h in air with a following regrinding and, in a second step, at 1025°C for 48 h in air. X-ray diffraction measurements were performed with a Seifert 3003 TT powder diffractometer using Cu-$K_\alpha$ radiation. All patterns could be well refined by the Rietveld method in space group Im-3 with R-values below 5%, proving phase purity and being consistent with literature [17]. For the dielectric measurements, the reground powders were pressed with 0.8 GPa into discs with diameters of 6 mm and typical thicknesses of 1 mm. All samples were sintered at 1050°C for 3 hours. The grain sizes were determined by scanning electron microscopy (SEM). After sintering, silver paint or sputtered gold contacts (thickness 200 nm) were applied to opposite sides of the samples. The complex conductivity $\sigma^* = \sigma' + i\sigma''$ and permittivity $\varepsilon^* = \varepsilon' - i\varepsilon''$ were measured over a frequency range of up to eight decades as detailed in Refs. 7 and 19. The applied ac voltage was 1 V at frequencies $\nu < 1$ MHz and 400 mV for $\nu > 1$ MHz.

## 3. Results and discussion

Figure 1 shows $\varepsilon'(T)$ of all investigated materials at 1.16 kHz. In all cases $\varepsilon'(T)$ tends to exhibit a plateau at high temperatures with only weakly temperature-dependent values ranging from $\varepsilon'_{giant} \approx 250$ ($Ln$ = Tm, Ho) up to 2000 ($Ln$ = Ce, La, Pr, Dy). Thus for all samples large values of $\varepsilon'$ are revealed and with $\varepsilon' > 1000$, especially for the La-, Ce-, Pr- and Dy-compounds the dielectric constant assumes values as also often found in CCTO ceramics [7,9,12,20]. With decreasing temperature, a step-like decrease of $\varepsilon'(T)$ shows up in all materials. It shifts to higher temperatures with increasing frequency (not shown), which reflects typical relaxational behaviour. A similar relaxation is also found in CCTO and is nowadays generally accepted as being of Maxwell-Wagner (MW) type. MW relaxations are non-intrinsic and usually can be described by an equivalent circuit consisting of the bulk contribution, connected in series to a parallel RC-circuit with $R$ and $C$ being much higher than the corresponding bulk quantities [6]. Such high-resistance regions in the sample can arise, e.g., from insulating depletion layers at the electrode-sample interface, grain boundaries or other types of internal barriers. At low frequencies and/or high temperatures, the high capacitance of the thin insulating layers leads to the detection of giant values of the dielectric-constant. At high frequencies and/or low temperatures, the layer capacitance becomes shortened and the intrinsic bulk behaviour is detected [6].

As revealed by Fig. 1, the La-, Ce-, Pr- and Dy-containing LnCTO compounds show the highest values of $\varepsilon'_{giant}$ of all investigated materials. For $Ln$ = La and Pr, the relaxation step (Fig. 1) shows up at the lowest temperatures. In general, relaxation steps occur in both, $\varepsilon'(T)$ and $\varepsilon'(\nu)$ when the condition $\omega\tau = 1$ is fulfilled, with $\omega = 2\pi\nu$ and $\tau$ the relaxation time. Taking into account that $\tau(T)$ usually increases with decreasing temperature, for the La and Pr compounds the smallest values of $\tau$ can be assumed and thus their CDCs are expected to persist up to the highest frequencies. La$_{2/3}$Cu$_3$Ti$_4$O$_{12}$ was already investigated earlier [21] and thus in the following we focus on PCTO.

As demonstrated by the SEM picture presented in the inset of Fig. 2, the grain sizes of the polycrystalline PCTO sample vary from 2 to 6 µm, comparable to CCTO ceramics prepared in a similar way [12]. The main frames of Fig. 2 show the temperature dependences of $\varepsilon'$ (a) and $\sigma'$ (b) of PCTO for various frequencies. In $\varepsilon'(T)$, the typical frequency-dependent shift of the relaxation steps is observed [6]. Correspondingly, peaks show up in $\sigma'(T)$ and, thus, also in $\varepsilon''(T)$ (both quantities are connected via $\varepsilon'' \propto \sigma'/\nu$). For PCTO, we find $\varepsilon'_{intrinsic} \approx 65$, which is lower than in CCTO ($\varepsilon'_{intrinsic} \approx 85$) [12]. This reduction points to a significantly stiffer phonon spectrum or phonons with lower polar weight, a fact that deserves further investigation.

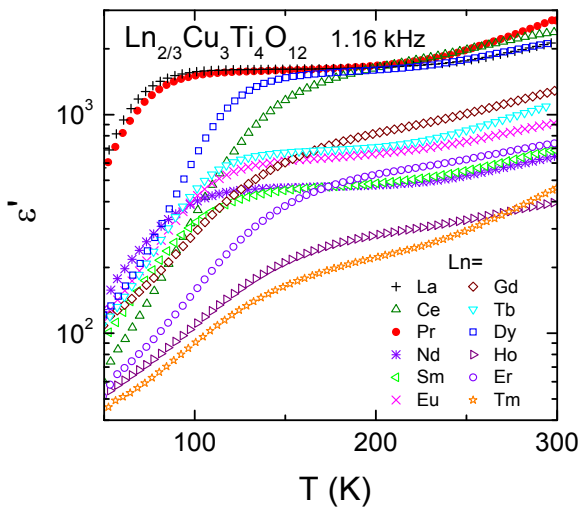

**Fig. 1.** (Colour online) Temperature-dependent dielectric constants of the investigated LnCTO-compounds at 1.16 kHz (silver-paint contacts).

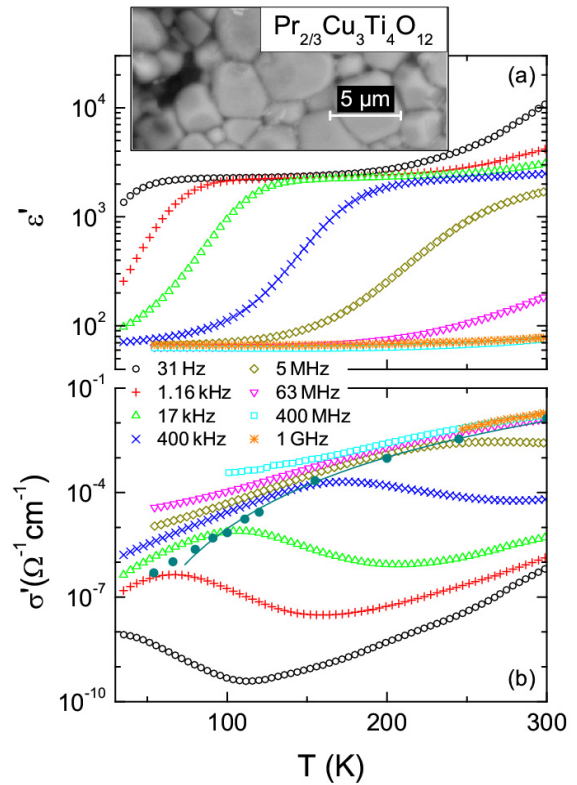

**Fig. 2.** (Colour online) Temperature-dependent $\varepsilon'$ (a) and $\sigma'$ (b) of PCTO at various frequencies using sputtered gold contacts. The filled circles in (b) represent the dc conductivity as determined from fits of the spectra of Fig. 4. The line is a fit with the VRH model with $T_0 = 7.8 \times 10^7$ K. The inset shows an SEM image, revealing grain sizes of 2 - 6 µm.

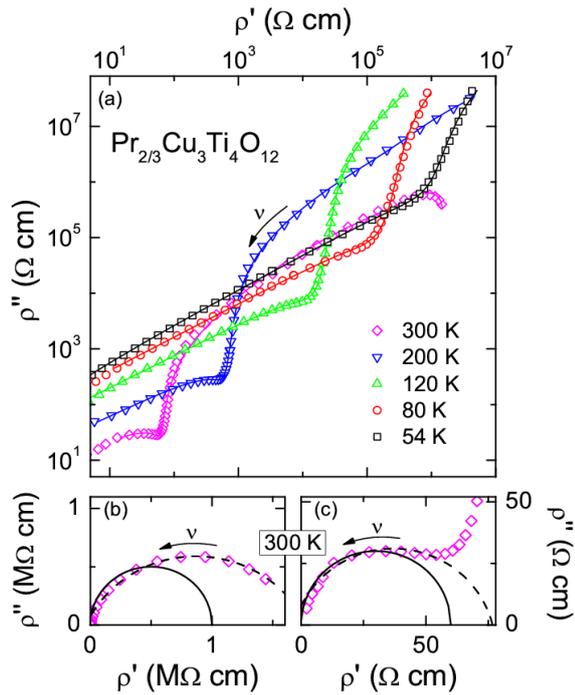

**Fig. 3.** (Colour online) Complex impedance-plane plot of the frequency-dependent resistivity of PCTO. In (a), showing the results for various temperatures, a double-logarithmic scale is used. The lines have been calculated from the fits curves in Fig. 4. In (b) the low- and in (c) the high-frequency part of the curve at 300 K are shown using a linear scale. The solid lines are semicircles whose centres are fixed to $\rho'' = 0$. The centres of the semicircles shown by the dashed lines are "sunken" below the abscissa. The arrows in (a) - (c) indicate increasing frequencies.

To elucidate the origin of the observed colossal values of $\varepsilon'$, in Fig. 3 we show our frequency-dependent results in the time-honoured complex impedance-plane representation. In these plots grain-boundaries or other barrier layer contributions should lead to separate semicircles, in addition to the one caused by the bulk. As our broadband data cover a huge range of impedance values, the double-logarithmic representation of Fig. 3(a) is used to provide a first overview. Two different contributions, separated by a kink, can be clearly distinguished for all temperatures. As an example, Figs. 3(b) and (c) show the low- and high-frequency features, respectively, at 300 K in linear representation. The observed succession of two approximate semicircles clearly points to non-intrinsic barrier-layer polarization governing the low-frequency behaviour and thus generating the colossal dielectric constants. As demonstrated by the lines in (b), only a "sunken" semicircle can describe the low-frequency data, which may point to a distribution of relaxation times [22]. Alternatively, two semicircles could be used instead. The high-frequency semicircle in Fig. 3(c) can be ascribed to the intrinsic bulk contribution. An extrapolation to zero frequency reveals an intrinsic dc resistivity of the order of $10^2$ $\Omega$cm. A precise determination is hampered by the deviations from a perfect semicircle. This may be due to the overlap with the low-frequency semicircle but may also point to an intrinsic frequency dependence of the conductivity [22]. It should be clearly noted that the impedance-plane plots do not provide information on the nature of the barrier layers that cause the low-frequency semicircle and the colossal epsilon values. Any effect that can be modelled by a parallel RC circuit (e.g., grain boundaries or electrode depletion layers) will lead to such an additional semicircle.

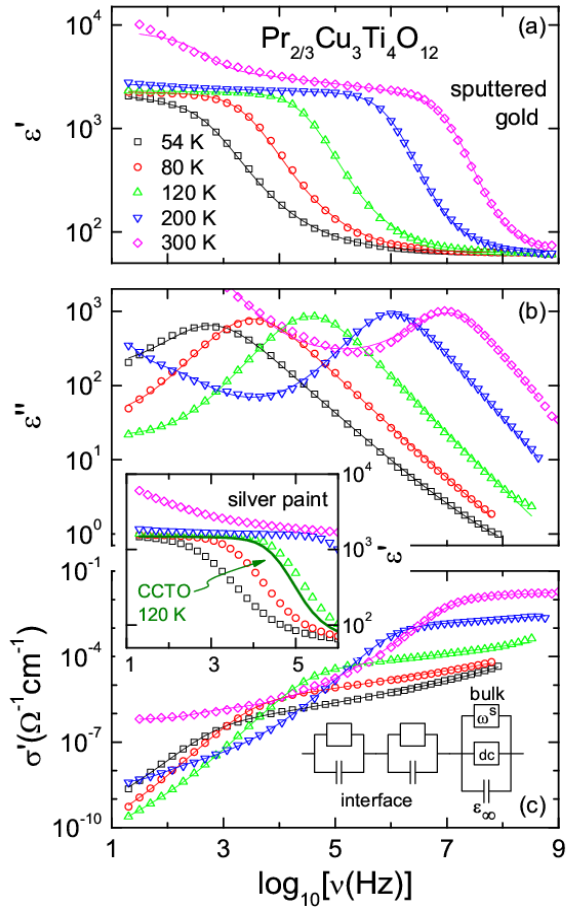

**Fig. 4.** (Colour online) Frequency-dependent $\varepsilon'$ (a), $\varepsilon''$ (b) and $\sigma'$ (c) of PCTO at various temperatures. The lines are fits using the equivalent circuit indicated in (c). The inset shows $\varepsilon'$ for the same sample with silver paint contacts. The line is the result for 3 h tempered CCTO measured at 120 K with silver-paint contacts (from [12]).

While impedance plots are suitable for a quick overview, a more sophisticated analysis can be provided by performing simultaneous fits of the real and imaginary part of the dielectric constant. This is done in Fig. 4 showing broadband dielectric spectra of PCTO for various temperatures. The MW relaxation is the dominant feature with well-pronounced, strongly temperature-dependent steps in $\varepsilon'(\nu)$ [Fig. 4(a)] and the corresponding peaks in the loss (b). In addition, at room temperature and low frequencies ($\nu < 1$ kHz) a further increase of $\varepsilon'(\nu)$ is observed where the dielectric constant reaches even higher values up to $10^4$ [Fig. 4(a)]. This may evidence a second relaxation, which is also known to appear in CCTO [7,11,12,23]. The data of Fig. 4 can be fitted using the equivalent circuit shown in Fig. 4(c). This circuit is described in detail, e.g., in [12] where it was demonstrated to provide a good description of the spectra of CCTO. The two RC circuits connected in series to the bulk elements account for the two observed MW contributions, e.g., caused by electrode polarization and any kind of IBLC. Below room temperature, only a single interface-related RC-element is sufficient to fit the data. As demonstrated by the lines in Fig. 4, this model also works well for PCTO. Also the results in the impedance plane are well described in this way [lines in Fig. 3(a)].

The fits provide information on the intrinsic dc conductivity $\sigma_{dc}$, whose temperature dependence is included in Fig. 2(b) (solid circles). It corresponds to the high-frequency plateau value in $\sigma'(\nu)$ as seen in Fig. 4(c) [6]. The dc resistivity $1/\sigma_{dc}$ agrees reasonably with the bulk dc resistivities read off in the impedance plots [see, e.g., Fig. 3(c)]. Its temperature dependence can be rather well described by $\sigma_{dc} \sim \exp[-(T_0/T)^{1/4}]$ [line in Fig. 2(b)], which is typical for variable range hopping (VRH) [24]. Thus $\sigma_{dc}$ in PCTO reveals similar characteristics as for CCTO but its absolute values are by about a factor of ten lower [cf., e.g., Figs. 1(b) and 2 in Ref. 12]. Therefore, in PCTO the intrinsic dielectric loss, which in both materials is completely determined by charge transport, is smaller than in CCTO, an important point for any application. However, its room temperature value still needs to be reduced for a straightforward application as dielectric material. The conductvity in this class of materials most likely results from oxygen non-stoichiometry [3]. Recently it has been shown that the conductivity of CCTO can be drastically reduced alread by marginal doping [25], a possible route also for PCTO. The equivalent circuit contains an additional element assuming a contribution $\sigma' \sim \nu^s$ with $s < 1$. This ac contribution accounts for the increase of $\sigma'$ with frequency observed, e.g., at 54 K and $\nu > 10^5$ Hz in Fig. 4(c). It is typical for hopping conductivity [26] and consistent with the VRH found from the dc conductivity [24,26]. Similar behaviour was also reported for CCTO [12]. It is this power law that leads to the sunken semicircle in Fig. 3(c) [22].

There is an ongoing discussion concerning the origin of the CDC in CCTO: Among the discussed mechanisms are surface depletion-layers, due to the formation of MIS or Schottky diodes [6,7,11,12,13], or internal layers like grain boundaries [3,4,8,9,10]. To obtain hints on the prevailing mechanism in PCTO, measurements with different types of metal contacts were performed. The inset of Fig. 4 shows $\varepsilon'(\nu)$ measured with silver-paint contacts, which closely resembles the corresponding curves with sputtered gold contacts in Fig. 4(a). Quite in contrast to CCTO [7,11,12] and various other materials with CDCs [15,27], using different electrode materials obviously has no significant impact on $\varepsilon'$ of PCTO. Thus, an IBLC mechanism, caused by insulating grain boundaries, is the most likely model. Here the surfaces of the grains are assumed to be insulating, which can arise, e.g., by deviations of the oxygen stoichiometry from that of the semiconducting bulk [3]. The remaining question is: Does grain growth result in a higher dielectric constant, just as observed in CCTO [9,12,23]? However, we found that in PCTO, in contrast to CCTO, larger grains cannot be easily attained by standard tempering procedures and further investigations are necessary to answer this question. In any case, it is clear that the dielectric constant in PCTO closely follows the behaviour observed in CCTO. This is further evidenced by the inset of Fig. 4, where $\varepsilon'(\nu)$ at 120 K of a CCTO ceramic with similar grain size, measured using silver-paint contacts, is included for comparison. The $\varepsilon'(\nu)$ curves of both materials are nearly identical.

## 4. Summary

In summary, we have performed dielectric spectroscopy in a broad frequency range on various, until now not or only rarely investigated LnCTO samples. In some of these compounds we have detected CDCs, similar to CCTO ceramics. Among the investigated compounds, PCTO, which has proved a promising material with dielectric properties comparable to those of CCTO, is investigated in more detail. Just as for CCTO [3], plots in the complex impedance plane reveal clear evidence for barrier-layer polarization causing the CDCs. Using different contact materials results in no significant change in the dielectric constant and thus an IBLC mechanism, most likely arising from grain boundaries, seems to prevail. The temperature and frequency dependence of the intrinsic conductivity of PCTO points to VRH as the dominating charge-transport process. With further optimization, especially by achieving larger grain sizes in polycrystalline samples or by the preparation of single crystals, dielectric constants in excess of $10^4$ should be within reach in PCTO. Overall, it seems that there is nothing special in CCTO that would make it stand out from other, isostructural compounds. The investigation of further members of this large group of materials should reveal more CDC materials that may exhibit properties even better suited for application.

## Acknowledgments

This work was supported by the Commission of the European Communities, STREP: NUOTO, NMP3-CT-2006-032644 and by the DFG via the SFB 484. We thank P. Steinbeiß for preparation assistance and R. Merkle from the Anwenderzentrum Material- und Umweltforschung Augsburg for technical support in the ESEM measurements.